\documentclass[preprint]{revtex4}
\usepackage{graphicx}
\usepackage{color}

\begin{document}
\title{Non-Hermitian delocalization and the extinction transition.}
\author{ David A. Kessler and  Nadav M.  Shnerb
 } \affiliation{Physics Department, Bar Ilan University, Ramat-Gan
52900, Israel
 }

\begin{abstract}
Logistic growth on a static heterogenous substrate is studied both
above and below the drift-induced delocalization transition. Using
stochastic, agent-based simulations the delocalization of the
highest eigenfunction is connected with the large $N$ limit of the
stochastic theory, as the localization length of the deterministic
theory controls the divergence of the spatial  correlation length at
the transition. Any finite colony made  of discrete agents  is
washed away from  a    heterogeneity with compact support in the
presence of strong  wind, thus the transition belongs to the
directed percolation universality class.  Some of the difficulties
in the analysis of the  extinction transition in the presence of a
localized active state are discussed.

\end{abstract}


\maketitle

The effect of drift on inhomogeneous systems that exhibit growth and
propagation has attracted much interest in the last decade
\cite{efetov, hatano,pik,been,lev,nel,karin,karin1,ind,pie,lin}.
When the time evolution of a system is governed by a real symmetric
evolution operator it may support both extended and localized
eigenstates. The eigenstates of a quantum particle in a single
potential well, for example, are either localized inside the wall or
extended above some threshold energy. In the presence of drift, or
other non-Hermitian perturbation~\cite{efetov}, the system undergoes
a phase transition where localized wavefunctions become extended,
and the corresponding eigenvalues migrate from the real axis to the
complex plane. This transition was first analyzed by Hatano and
Nelson~\cite{hatano} in the context of flux lines in high $T_c$
superconductors with columnar defects subjected to a tilted external
magnetic field. Since then, many authors have considered this
transition in different fields, e.g. hydrodynamics~\cite{pik},
random lasers~\cite{been}, and quantum dots~\cite{lev} among many
others.

Of particular interest, both theoretically~\cite{nel, karin,
karin1,ind} and experimentally~\cite{pie,lin},  is the
delocalization transition for bacterial colonies on a heterogeneous
substrate in the presence of drift. A logistic growth of a motile
population on a 1d static spatially heterogenous substrate is
described by:
\begin{equation} \label{deter}
\frac{\partial c(x,t)}{\partial t} = D \nabla^2 c(x,t) + v
\frac{\partial c(x,t)}{\partial x}+ a(x) c(x,t) -  c^2(x,t).
\end{equation}
In the absence of drift term ($v=0$) and for a homogenous
environment ($a =a_0 \equiv \sigma - \mu$, where $a_0$, the
difference between  the birth rate $\sigma$ and the death rate
$\mu$, is independent of spatial location \cite{us}) one gets the
celebrated Fisher-Kolomogorov-Petrovsky-Piscounov equation (FKPP), a
generic description of an invasion of a stable state $(c^* = a_0)$
into an unstable one $c^* = 0$. In the asymptotic long-time limit
this system supports a front that travels with constant speed $v_F =
2 \sqrt{D a_0}$. In the homogenous case the eigenstates of the
linearized evolution operator
\begin{equation}
{\cal{L}}= D \nabla^2 c(x,t) + v \frac{\partial c(x,t)}{\partial x}+
a_0 c(x,t).
\end{equation}
are  extended sinusoidal functions  and the drift corresponds to a
simple Galilean transformation.

Things change when  translational invariance is broken, i.e.,
in the presence of spatial inhomogeneity. Two main types  of
heterogenous growth are  considered in the literature
\cite{nel,karin,karin1,pie,lin}: A "single oasis" case, where the
growth rate is larger on a spatial domain, and the disordered case,
where $a(x) = a_0 + \delta a(x)$,  $\delta a$ being taken from some
random distribution with zero mean. In both cases the spectrum of
the linear evolution operator admits localized wavefunctions; if $\{
\phi^0_n(x), \Gamma_n \}$ is the set of eigenfunctions and the
corresponding eigenvalues of   ${\cal L}(v=0)$, at least some
eigenstates in the tail of the spectrum (or all the states for a
disordered potential below 2d) are exponentially localized. The
effect of small drift on a localized eigenstates is trivial:
\begin{equation} \label{gauge}
\phi^v_n(x) = e^{vx/2D} \phi^0_n(x) \quad,\quad \Gamma^v_n = \Gamma^0_n
- \frac{v^2}{4D}.
\end{equation}
This "gauge invariance" breaks down at $v^c_n = 2D/\xi_n$, where
$\xi_n$ is the localization length of the $n$th  eigenstate. Above
$v^c_n$ the eigenstate delocalizes  and the boundary conditions
begin to play an important role: e.g., for periodic boundary
conditions the eigenvalues that correspond to delocalized
eigenstates become complex~\cite{prl}. The spectrum then takes the
form of a "bubble" in the complex plane, where the localized
eigenstates correspond to the spectral points in the tail, since the
localization length in the center of the band is larger. A
non-Hermitian "mobility edge" appears between the two regimes.
Increasing $v$ even more, the bubble spreads and captures more and
more spectral points, and at the end the ground state also
delocalizes. The Perron-Frobenius Theorem \cite{pp} ensures that the
highest eigenstate stays on the real line, and the delocalization
transition is identified by the breakdown of the trivial gauge (Eq.
\ref{gauge}) and the vanishing of the spectral gap~\cite{nel,prl}.

Figure 1 shows some examples of the spectrum of ${\cal L}$, together
with a sketch of the phase diagram, for the single oasis scenario.
In the absence of drift there is a single localized state at the
right edge of the spectrum (if the oasis is large, a few localized
states exist), followed by a continuum of states that correspond to
extended eigenfunctions. Even a small drift is enough to push the
delocalized eigenvalues to the complex plane, but the localized
state only develops a slight asymmetry with almost no effect on the
eigenvalue. Only for high enough drift does the highest eigenstate
delocalize and the gap disappear.   A change of $a_0$ corresponds to
a rigid shift of the whole spectrum along the real line.   Thus,
three regimes exist in the drift-proliferation parameter space: the
extinction region, where the real part of all the eigenvalues is
negative; the localized region, where only the localized states
admit positive growth rate; and the proliferation regime, where both
localized and extended states may grow. Above $v^c_0$ only the
extinction and the proliferation regions exist.

\begin{figure*}
\includegraphics[width=13cm]{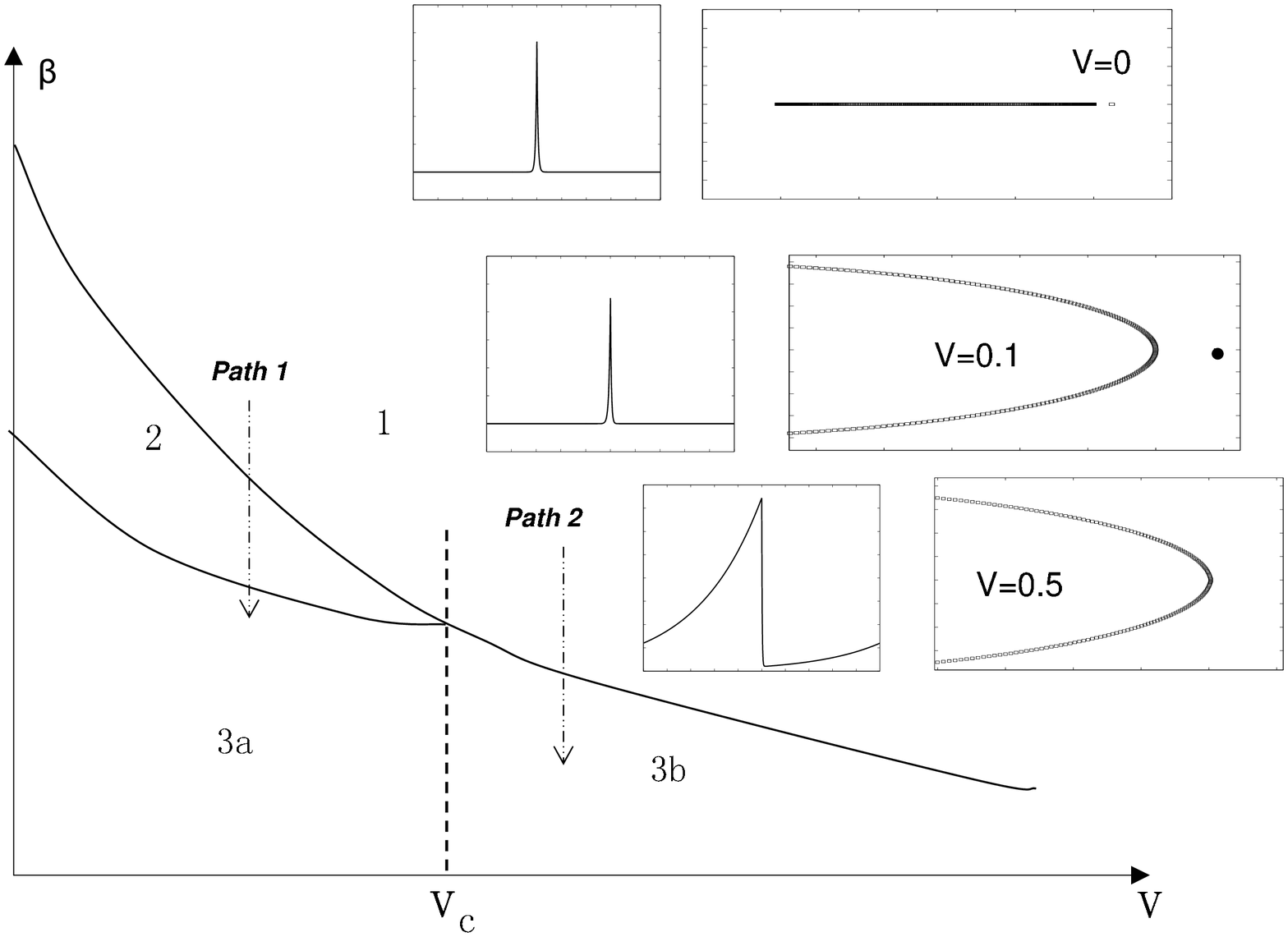}
 \caption{A sketch of the "phase diagram" for a single oasis in the
death-rate - drift space. Region (1) is the extinction region
 where the highest eigenvalue is negative. In region (2) only  the localized
state admits an  eigenvalues with positive real part, and in regions
(3a) and (3b) extended eigenstates become "active".
 In the right side of the figure the highest eigenstate and the spectrum
in the complex plane are plotted for the cases of no drift ($v=0$),
small drift ($v = 0.1$) and large drift ($v=0.5$) for a single
 oasis.}
 \label{fig1}
\end{figure*}

The above discussion is, however, too naive. Bacterial systems
are not deterministic, and are composed of discrete objects that may
die, reproduce or migrate with some probability that depends on the
local environmental conditions. The bacterial population at a
certain point is not a deterministically varying continuous
quantity, like $c(x)$, but a discrete number that undergoes
stochastic processes, e.g.,  $A \to 2A, \ A+A \to \emptyset , \  A
\to \emptyset $ etc. Like in many other branches of science, the
deterministic dynamics is an approximate description of the system
that becomes exact where the effect of stochasticity vanishes. In
the case considered here the demographic stochasticity becomes
negligible when the density of agents is large, since the relative
fluctuations scale with $1/\sqrt{N}$. Technically, the exact
stochastic Master equation is replaced by a deterministic
description using the Kramers-Moyal expansion, or more rigorously by
van-Kempen's $\Omega$ expansion  and related methods
\cite{vk,gardiner}.  Joo and Lebowitz \cite{joo} have already
pointed out that in the limit of large $N$ one should expect a
population density distribution that follows the spatial features of
the active eigenstates, i.e., the eigenstates for which $Re(\Gamma)
> 0$. Here, on the other hand, we want to discuss the effects of
spatial  heterogeneity and drift for a dilute system; i.e.,
\emph{close to the extinction transition}. In that case the
$1/\sqrt{N}$ expansions are invalid and so we resort to numerical simulations.

Let us first present some general considerations. Grassberger and
Janssen~\cite{jan}  suggested long ago that the extinction
transition to a single absorbing state on a homogenous substrate
belongs (in the absence of special additional symmetries) to the
directed percolation (DP) equivalence class, independent of the
microscopic details of the stochastic process. DP is a continuous
transition and the correlation length and  correlation times diverge
at the transition point with their characteristic exponents (see
\cite{1} for a general review). On a homogenous substrate the
correlation length is the only length scale of the problem.  On a
static heterogenous substrate, on the other hand, another length
scale appears - the localization length. How do these two quantities
relate to each other? What are the properties of the stochastic
extinction transition below and above the deterministic
delocalization transition? In what sense is Eq. (\ref{deter})  a
deterministic limit of a stochastic process when the effect of
stochasticity is important; i.e., close to the extinction
transition?

Recently, this last question has been addressed  for the transition
on a homogenous substrate \cite{kessler}. It turns out that the
transition is always in the DP equivalence class, but the carrying
capacity of the system, N, determines the location of the transition
and, more important, the width of the transition zone. In the
deterministic theory the correlation length is zero both below and
above the  transition (any initial density fluctuation  simply
decays exponentially to the stable state   and its spread during
this process is negligible). The spatial correlation length for the
stochastic process satisfies $\xi_\perp \sim
\Delta^{-\nu_\perp}/N^\kappa$, where $\Delta$ is the distance from
the transition. If the system parameters are such that its
deterministic analogue is at the transition point, then $\Delta(N)
\sim N^{-\tau}$ and $\Delta$ vanishes at the deterministic limit.
Under these conditions $\xi_\perp \sim N^\gamma$, where $\gamma =
\tau \nu_\perp - \kappa$.  This implies that for any \emph{finite}
$\Delta$ for large enough $N$ the correlation length shrinks to zero
and the deterministic description holds, whereas for any finite $N$,
for small enough $\Delta$ the system enters the transition zone and
the deterministic description collapses. Both  $\kappa$ and $\tau$
depend on the deterministic features of the model as explained in
\cite{kessler}, but in any case $\gamma>0$ so the deterministic
limit never exists at the transition point.

In order to simulate the heterogenous system in the large $N$ limit,
an individual based  model that allows for an accurate determination
of the transition point in the $N \to \infty$ limit is used. We
consider a logistic growth process on a one dimensional lattice with
periodic boundary conditions; Euler integration is used with small,
but finite, $\Delta t$. The number if agents at the $i$-th lattice
site is an integer $n_i$, and each cycle of the  Monte-Carlo
simulation involves two consecutive  steps. The first step is the
reaction: each of the agents at the site produces an offspring with
probability $(\sigma_0 + \delta \sigma_i) (1-n_i/N_0) \Delta t$, and
dies with probability $\mu \  \Delta t$. In the second, diffusion
step, any agent is selected for migration with probability $2 \chi
\Delta t$, then chooses its destination - to the left with
probability $q_L = (1+\nu)/2 $ or to the right with probability
$1-q_L$. To avoid artificial drift as a result of the sequential
update of lattice sites, parallel update was used; $n_i$ is updated
only after the diffusion cycle is completed.

In the linearized deterministic limit this model corresponds to an
$L$ dimensional map, where $L$ is the number of sites. This map is
given by the multiplication of the reaction matrix, $R_{i,j} =
\delta_{i,j} [1+ \Delta t (\sigma_0 + \delta \sigma_i - \mu)]$, by
the diffusion matrix that takes the form (up to the boundary
conditions) $ D_{i,j} = \delta_{i,j} (1- 2 \chi \Delta t) +
\delta_{i,j+1} \chi (1+\nu) \Delta t + \delta_{i+1,j} \chi (1-\nu)
\Delta t$. Diagonalizing the product $D R$ one finds the highest
eigenvalue $\tilde{\Gamma}_0$  and the corresponding eigenvector,
$\phi_0^i$; adding another death process, where each particle in the
MC simulation is selected to die after any cycle with probability
$1/\\tilde{Gamma}_0$, ensures that the system is \emph{exactly} at
the transition point for $N_0 \to \infty$. In different words, the
agent-based system is simulated with a parameter set that ensures
$\Gamma_0 = 0$  in the deterministic limit.

Clearly, a system with  finite carrying capacity $N_0$ is  always
closer to extinction than the deterministic system when all other
parameters are equal. This implies that, scaling the parameters as
described above and  increasing $N_0$, the system is always in the
extinction phase and reaches the transition exactly at $N_0 =
\infty$. In Fig. \ref{fig2}, $\xi_\perp$, the correlation length, is
plotted vs. $N_0$ on a log-log scale and reveals the real meaning of
the deterministic delocalization transition:  below $v_c$, i.e.,
when $\phi_0$ is localized, the correlation length associated with
the stochastic process first grows and then saturates to the
deterministic localization length $ \xi / [1-v \xi /(2D)]$. This
demonstrates the fact that the state that becomes active at the
transition is \emph{localized} and the correlation length of the
stochastic process can not grow beyond this deterministic length.

\begin{figure}
 \includegraphics[width=8cm]{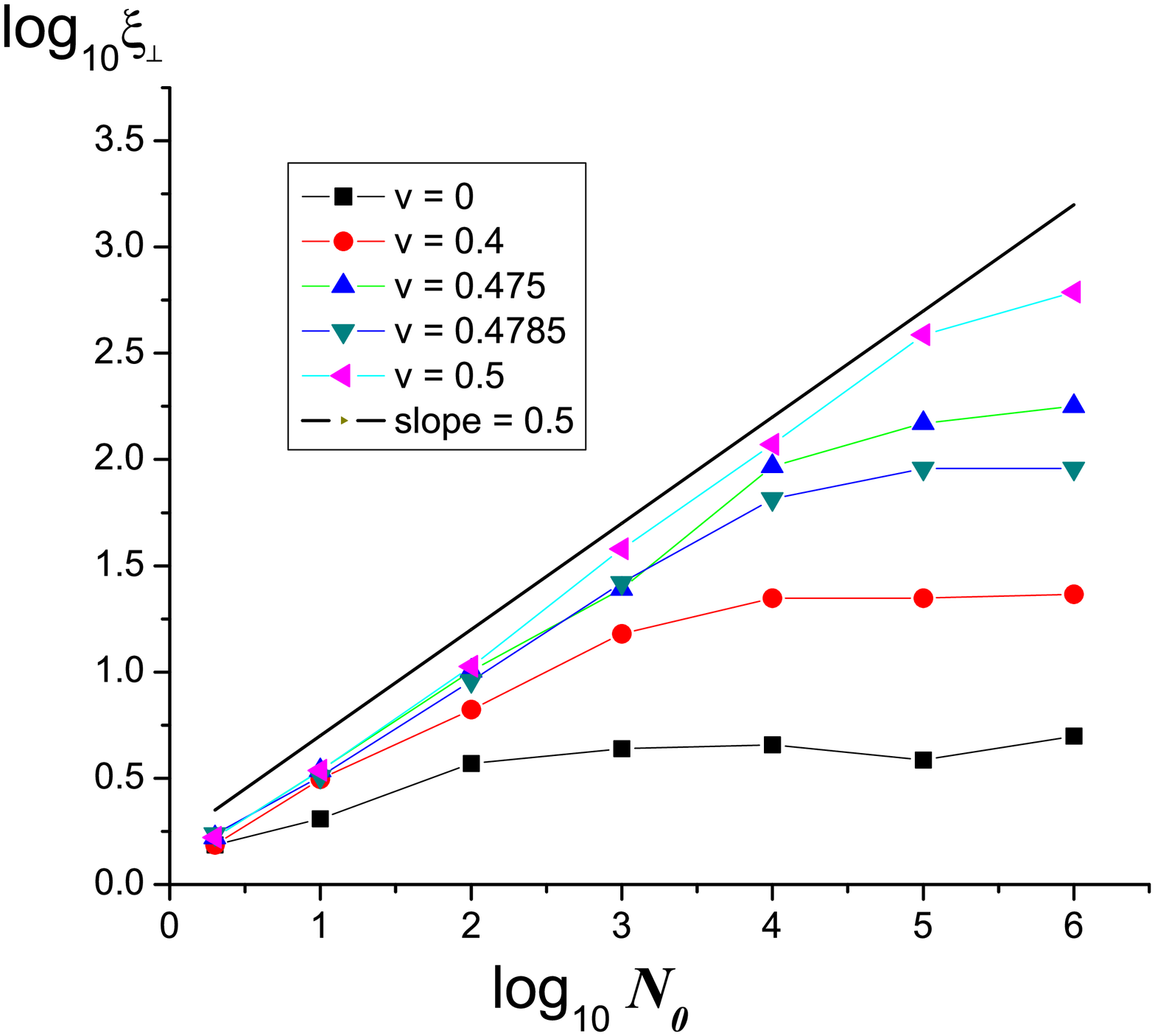}
 \caption{The log of the transverse correlation length $\xi_\perp$
 (in the direction of the drift) vs. $log(N_0)$ for a
system of length $L=2000$. Here $\xi = 0.2, \sigma_0=\mu =1$ and the
heterogeneity $\delta \sigma_i = 0.2 \delta_{i,L/2}$.  }
 \label{fig2}
\end{figure}

On the other hand, above $v_c$ the correlation length grows
unboundedly with $N_0$. The data is consistent with  $\xi_\perp \sim
N_0^{0.5}$, which is  (up to logarithmic corrections), the lifetime
of a well-mixed system at the deterministic transition point
\cite{san}. This reflects the fact that the delocalized system is
not really one dimensional but rather $"0+1"$ dimensional, with the
spatial direction playing the role of time. Life in that system is a
result of a drift from a source, not of uniform growth, and the
lifetime of the colony at the oasis determines the spatial extent
reached by its decedents.

For finite $N_0$ and with constant drift velocity $v$, the system
undergoes undergone an extinction transition  as $a_0$ decreases.
This may happen either via the localized phase (e.g., along  Path 1
shown by the  arrow in Figure 1), or directly to the delocalized
phase (Path 2 in figure 1). While for $N_0 \to \infty$ the
transition happens when $Re(\Gamma_0)$ touches zero, for any finite
$N_0$ the transition takes place when a finite region of the upper
part of the spectrum is above zero (inset of Fig. 3). For a single
oasis (or otherwise when the number of oases is finite) all the
localized states decay in the long run as a result of demographic
stochasticity; only when the extended eigenstates are "excited"
(their eigenvalues cross to the positive real part of the spectrum)
will the system be in its active phase. As a result, the scenarios 1
and 2 can be seen to differ significantly.

Let us first consider path 2. Intuitively, above $v_c$ the colony is
carried off the oasis by the wind, thus the large-scale properties
of the system are identical with a homogenous substrate with drift
in the thermodynamic limit. More precisely, for finite $N_0$ the
transition occurs when a finite  part of the spectrum, made of
delocalized states, is already "excited" (i.e., $Re(\Gamma)>0$ for
the these eigenstates). Thus, there are two regimes. Deep in the
extinction phase all states decay, $Re(\Gamma)<0$.  The bacterial
density in this regime satisfies the deterministic solution $c(x,t)
= \exp(- |a_0| t - (x-x_0-vt)^2/4Dt)/\sqrt{4 \pi D t}$, where $x_0$
is the nucleation point. The overall occupation of a point, $C_T(x)
\equiv \int c(x,t) \ dt$, is thus a monotonically decreasing
function of $x-x_0$, with an exponential decay of the tail $C_T(x)
\sim exp(-(x-x_0)/\xi)$, where the localization length $\xi$ scales
like $D/(\sqrt{v^2 + 2 D|a_0|}-v)$.

Close to the transition point for finite $N_0$, on the other hand,
many linear states are already excited and the growth of the colony
is unaffected by the nonlinear competition at short times. Only
after the characteristic time $\xi_\|$  does nonlinearity suppress
the growth,  leading to extinction. Within this growth period the
system behaves deterministically and a "Fisher front" starts to
invade the empty region. As the wind velocity is larger than the
Fisher velocity above $v_c$ \cite{nel}, the maximum of $C_T(x)$
moves in the direction of the wind, as demonstrated in Figure 3.
This second regime vanishes at the deterministic limit; accordingly,
the detachment of the peak from the nucleation point disappears upon
increasing  $N_0$.

\begin{figure}
 \includegraphics[width=8cm]{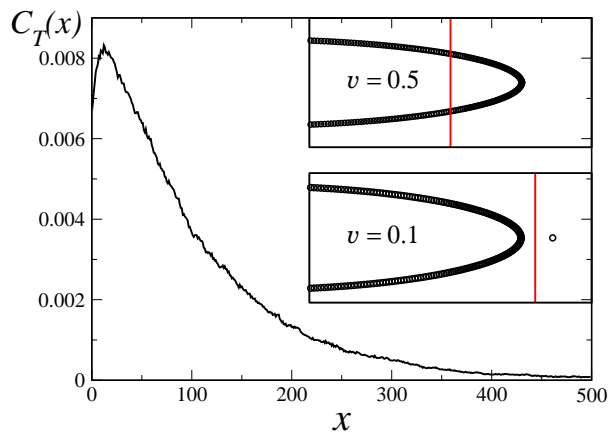}
 \caption{$C_T(x)$ vs. $x$ close to the transition ($\beta = 0.19$, $v=0.7$, other parameters identical with
those used in Fig. 2) for $N_0 =3$.
 The effect of the localized heterogeneity at $x=0$ is almost
 unseen. As emphasized in the upper inset, the transition takes
 place when many delocalized modes are also excited; as a result the
 system initially grows and the colony pushed to the  right by the
 wind, yielding a hump away from the oasis. This peak disappears for
 larger $\beta$ and for larger $N_0$ (results not shown). The lower
 inset exemplify the spectrum of the linearized evolution operator
 when the transition takes place along path 1 of Figure 1.}
 \label{fig2}
\end{figure}
The situation is completely different along path 1. The highest
state is now localized, and its nonlinear interaction differs
substantially from the interaction between extended states. If the
localization length is finite the oasis region decouples from the
rest of the system in the thermodynamic limit and the DP dynamics
happens in parallel with the zero dimensional stochastic process on
the oasis. This decoupling, however, is impossible at the bulk DP
transition, when $\xi_\bot$ diverges \cite{hi}. A related issue is
the transition in the presence of a finite density of randomly
distributed oases: below $v_c$ a nonuniversal Griffiths phase
appears between the active and the inactive parameter regions
\cite{ga}. In the deterministic limit only the highest localized
state becomes active at the transition, thus the Griffith phase
admits no deterministic limit, and its width shrinks to zero. These
last two observations suggest that the deterministic description of
the system by means of excited localized eigenstates is
insufficient, as the convergence of a finite $N$ system to the
deterministic limit is a very subtle issue, to be addressed in
subsequent publication.

\begin{acknowledgments}
 This work  of N.S. was supported by the EU 6th framework CO3 pathfinder.
\end{acknowledgments}


\begin{thebibliography}{}
\bibitem{efetov} K. B. Efetov, Phys. Rev. Lett. \textbf{79}, 491 (1997); Phys.
Rev. B \textbf{56}, 9630 (1997).
\bibitem{hatano} N. Hatano and D. R. Nelson, Phys. Rev. Lett. \textbf{77}, 570
(1996); Phys. Rev. B \textbf{56}, 8651 (1997).
\bibitem{pik} A. V. Straube and A. Pikovsky, Phys. Rev. Lett. \textbf{99}, 184503
(2007).
\bibitem{been} C. W. J. Beenakker, J. C. J. Paasschens, and P. W.
Brouwer, Phys. Rev. Lett. \textbf{76}, 1368 (1996).
\bibitem{lev} M. S. Rudner and L. S. Levitov, cond-mat/0807.2048
\bibitem{nel} D. R. Nelson and N. M. Shnerb, Phys. Rev. E \textbf{58}, 1383
(1998).
\bibitem{karin} K. A. Dahmen, D. R. Nelson, and N. M. Shnerb, J. Math.
Biol. \textbf{41}, 1
(2000); K. A. Dahmen, D. R. Nelson, and N. M. Shnerb, in \textit{Statistical
Mechanics of Biocomplexity}, edited by D. Reguera, J. M. G. Vilar,
and J. M. Rubi (Springer, Berlin, 1999).
\bibitem{karin1} A.R. Missel and K.A. Dahmen, Phys. Rev. Lett. \textbf{100},
058301 (2008).
\bibitem{ind} V. M. Kenkre and N. Kumar, cond-mat/0808.0172.
\bibitem{pie} T. Neicu, A. Pradhan, D. A. Larochelle, and A. Kudrolli,
Phys. Rev. E \textbf{6}2, 1059 (2000); N.M. Shnerb, Phys. Rev. E \textbf{63}, 011906
(2000).
\bibitem{lin} A.L. Lin, B.A. Mann, G. Torres-Oviedo, B.
Lincoln, J. K\"{a}s and H. L. Swinney, Biophysical Journal
\textbf{87}, 75-80 (2004).
\bibitem{us} Althogh in the deterministic limit, $\beta$ simply
shifts  $\alpha$, it is important to include an explicit death
term in the microscopic theory, as explained in \cite{kessler}.
\bibitem{prl} N. M. Shnerb and D. R. Nelson, Phys. Rev. Lett.
\textbf{80}, 5172 (1998).
\bibitem{jan} H. K. Janssen, Z. Phys. \textbf{B 42}, 151 (1981); P.
Grassberger, Z. Phys. \textbf{B 47}, 365 (1982).
\bibitem{pp}Bapat, R.B. and Raghavan, T.E.S., \textit{Nonnegative Matrices and
Applications, Encyclopedia of Mathematics and its Applications},
(Cambridge University Press, Cambridge, 1997).
\bibitem{joo} J. Joo and J.L. Lebowitz, Phys. Rev.  E \textbf{72}, 036112
(2005).
\bibitem{vk} N. G. van Kampen, \textit{Stochastic Processes in Physics and Chemistry, Third Edition}, (Elsevier, Amsterdam, 2007).
\bibitem{gardiner} C. W. Gardiner, \textit{Handbook of Stochastic Methods for Physics, Chemisty and the Natural Sciences, Third Edition}, (Springer-Verlag, Berlin, Heidelberg, New York, 2004).
\bibitem{1}H . Hinrichsen,
Adv.Phys. \textbf{49}, 815-958  (2000).
\bibitem{san} C. R. Doering, K. V. Sargsyan, and L. M. Sander, Multi-scale Model. Simul. \textbf{3}, 283 (2005).
\bibitem{kessler} D. Kessler and N.M. Shnerb, J. Phys. A \textbf{41}, 292003 (2008).
\bibitem{hi} See, in this regard,  A.C. Barato and H. Hinrichsen,  cond-mat/0802.3580.
\bibitem{ga} A.G. Moreira and R. Dickman, Phys. Rev. \textbf{E 54} R3090
(1996).

\end{thebibliography}
\end{document}